\documentclass[aps,prx,twocolumn,superscriptaddress,showpacs]{revtex4-1}
\usepackage{amsmath,amssymb}
\usepackage{graphicx}
\usepackage{natbib}
\usepackage{color}

\begin{document}
	
\title{Boosting the Terahertz Photoconductive Antenna Performance with Optimized Plasmonic Nanostructures}

\author{Sergey~Lepeshov}
\affiliation{ITMO University, St.Petersburg 197101, Russia}

\author{Andrei~Gorodetsky}
\affiliation{ITMO University, St.Petersburg 197101, Russia}
\affiliation{Aston Institute of Photonic Technologies, Aston University, Birmingham B4 7ET, UK}
\affiliation{Department of Engineering, Lancaster University, Lancaster LA1 4YW, UK, \\ and Cockcroft Institute of Accelerator Science, Warrington
WA4 4AD, UK}
\email{andrei@corp.ifmo.ru}

\author{Alexander~Krasnok}
\affiliation{Department of Electrical and Computer Engineering, The University of Texas at Austin, Austin, Texas 78712, USA}
\email{akrasnok@utexas.edu}

\author{Nikita~Toropov}
\affiliation{ITMO University, St.Petersburg 197101, Russia}

\author{Tigran~A.~Vartanyan}
\affiliation{ITMO University, St.Petersburg 197101, Russia}

\author{Pavel~Belov}
\affiliation{ITMO University, St.Petersburg 197101, Russia}

\author{Andrea~Al\'{u}}
\affiliation{Department of Electrical and Computer Engineering, The University of Texas at Austin, Austin, Texas 78712, USA}

\author{Edik~U.~Rafailov}
\affiliation{Aston Institute of Photonic Technologies, Aston University, Birmingham B4 7ET, UK}

\begin{abstract}
Advanced nanophotonics penetrates into other areas of science and technology, ranging from applied physics to biology and resulting in many fascinating cross-disciplinary applications. It has been recently demonstrated that suitably engineered light-matter interactions at the nanoscale can overcome the limitations of today's terahertz (THz) photoconductive antennas, making them one step closer to many practical implications. Here we push forward this concept by comprehensive numerical optimization and experimental investigation of a log-periodic THz photoconductive antenna coupled to a silver nanoantenna array. We shed light on the operation principles of the resulting hybrid THz antenna, providing an approach to boost its performance. By tailoring the size of silver nanoantennas and the distance between them, we obtain an enhancement of optical-to-THz conversion efficiency 2-fold larger compared with previously reported results, and the strongest enhancement is around 1~THz, a frequency range barely achievable by other compact THz sources. Moreover, we propose a cost-effective fabrication procedure to realize such hybrid THz antennas with optimized plasmonic nanostructures via thermal dewetting process, which does not require any post processing and makes the proposed solution very attractive for applications.
\end{abstract}

\maketitle

\begin{figure*}[!t]
\begin{center}
\includegraphics[width=0.99\linewidth]{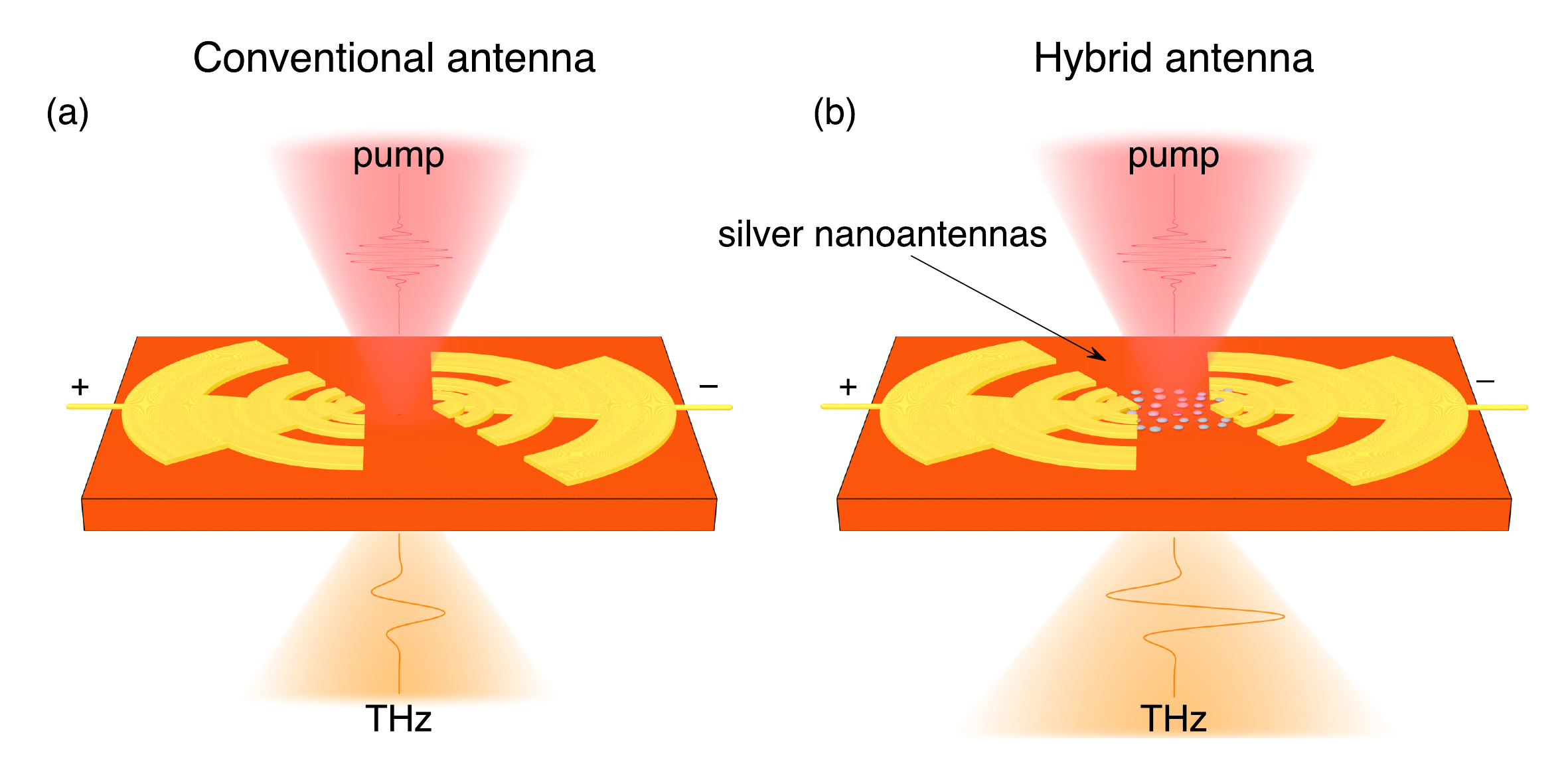}
\end{center}
\caption{Illustration of the conventional (a) and silver nanoantennas-loaded hybrid (b) photoconductive THz antenna.}\label{fig1}
\end{figure*}

Terahertz (THz) technology is now standing at the lab doorstep to real world applications. The THz spectral band of electromagnetic waves has found a wide range of perspective applications for spectroscopy~\cite{Schmuttenmaer2002, Jepsen2011,Jensen2013,Zhang2015}, microscopy~\cite{Hornett2016, Moon2015}, terahertz molecule sensing~\cite{Park2013,Smirnov2014, Yang2016b}, security imaging~\cite{Federici2005, Zhang2008}, detection of dangerous or illicit substances~\cite{Massaouti2013, Zhang2008, Zhou2008a} and ultrafast data transfer~\cite{Shams2015a, Nagatsuma2016b}. However, the widespread use of THz technologies is hampered by the absence of effective, compact and energy efficient THz sources operating at room temperatures. The most common source of \textit{coherent pulsed} THz radiation so far are the so-called THz photoconductive switches, or \textit{photoconductive antennas} (PCAs)~\cite{Auston84, Jeong_ACSN_12, Jooshesh2015, Lepeshov2016b}. A typical realization of such THz antenna in log-periodic design is presented in Figure~\ref{fig1}(a). Here, two (or more) conductive electrodes spaced by a gap are deposited onto a semiconductor surface. The electrodes are biased by an external voltage of several $V$, and the gap between them is pumped by femtosecond (fs) optical pulses. The principles of THz PCAs operation are based on the effect of ultrafast variations of surface photoconductivity of a semiconductor substrate under fs-laser irradiation: after exciting the gap between the electrodes with a fs-laser, the concentration of charge carriers increases sharply for a short period of time, and THz pulse generation occurs. Since the generated signal is a short single pulse hundreds of fs in duration, its spectrum spreads over several octaves in THz frequency range. The broad bandwidth is ideal for spectroscopic investigations of organic molecules~\cite{Fischer2002}, material science~\cite{Grischkowsky1990, Schmuttenmaer2002, Schmuttenmaer2004} as well as wireless THz transmitters and receivers~\cite{Akyildiz2014a}. However, conventional THz PCAs have a rather low conversion efficiency that prevents data transmission over long distances, they do not provide a large signal-to-noise ratio, and require ultrafast laser pump at wavelengths around $\lambda=$~800~nm, usually provided by expensive and bulky Ti:Sapphire lasers~\cite{Bespalov2008}. The low conversion efficiency is mainly related to two predominant factors: the photocarrier screening effect~\cite{Pedersen1993, Loata2007a}, and the low absorption coefficient of the surface layer of the semiconductor substrate. To overcome these limitations, optical nanoantennas~\cite{Krasnok2013, AluBook, Novotny2009} have been proposed to be placed in the gap of THz PCA~\cite{Lepeshov2016b,Chen2013,Berry2013,Heshmat2012,Jeong_ACSN_12}.

Optical nanoantennas are resonant nanostructures that are capable to transform incident optical waves into a strongly localized near-field~\cite{Alu2008, Alu2008b, Alu2010a, AluBook}. Nowadays, nanoantennas are commonly used to enhance the absorption coefficient of a semiconductor substrate of PCAs~\cite{Lepeshov2016b,Berry2013}. The nanoantenna-based PCAs have been called \textit{hybrid terahertz-optical PCAs}. The schematic presentation of such antenna is illustrated in Figure~\ref{fig1}(b). It has been shown that this solution provides high fs-laser pump absorption, shorter photocarrier lifetimes and excellent thermal efficiency~\cite{Lepeshov2016b}. Despite the fact that many nanoantenna designs have been studied for enhancement of the THz generation from PCA, an optimized geometry has not yet been proposed. Here we push forward this concept by comprehensive numerical optimization and experimental investigation of a log-periodic THz photoconductive antenna coupled to a silver (Ag) nanoantenna array arranged in the gap of a THz antenna. We shed light on the operation principles of the resulting hybrid THz antenna providing an approach to significantly boost its performance. By tailoring the size of silver nanoantennas and the distance between them, we obtain an enhancement of optical-to-THz conversion efficiency 2-fold larger in comparison with previously reported results. As a byproduct, we propose a \textit{cost-effective} fabrication procedure allowing to produce such hybrid THz antennas with optimized plasmonic nanostructures via thermal dewetting process.

\begin{figure*}[!t]
	\begin{center}
		\includegraphics[width=0.99\linewidth]{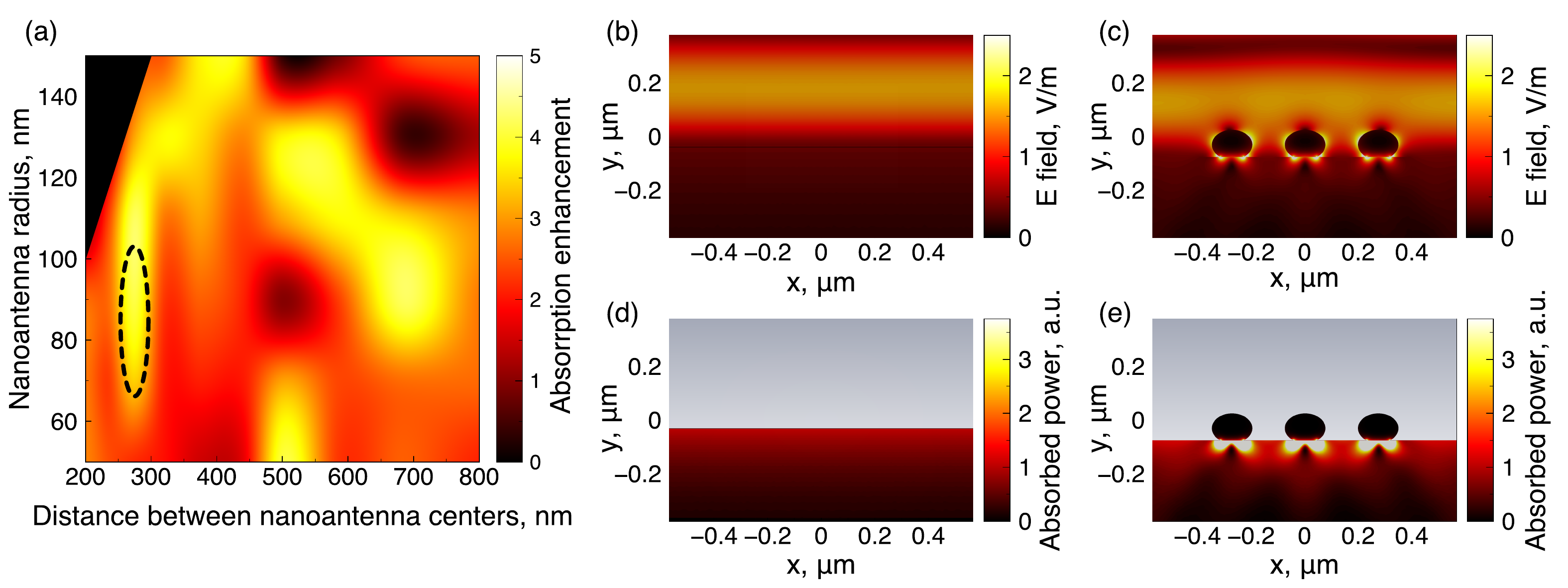}
	\end{center}
	\caption{(a)~Dependence of the absorption enhancement inside the GaAs substrate on the distance between nanoantennas and their radii (D/2). Electric field~(b,c) and density of absorbed power (d,e) distributions over the pure GaAs substrate and over the substrate containing Ag nanoantennas, correspondingly. The excitation plane wave (with the electric field strength of 1) propagates normally to the surface. The wavelength of excitation is 800~nm.}\label{fig2}
\end{figure*}

We start our analysis with numerical calculations of a spheroidal silver (Ag) nanoantenna array arranged over a high-index gallium arsenide (GaAs) semiconductor substrate. Nanoantennas have the form of an oblate spheroid with larger axis D and minor axis d, and these quantities are subject to optimization. It is well-known that such metallic (particularly Ag) nanoparticles provide localized plasmonic resonances in the optical range~\cite{Maier_book,LukasNovotny2012} that manifest themselves in a strong localization of the electric field near the nanoparticle due to conversion of the freely propagating incident waves into the near-field. If the size of the nanoparticles is sufficiently small (in terms of wavelength of the incident wave), the nanoparticle can be treated as an oscillating electric dipole. For the numerical simulations, CST~Microwave~Studio software package has been used. The calculated numerical results of the absorption enhancement at the wavelength of 800~nm as a function of the nanoantenna radius (D/2) and distance between nanoantenna centers (a) are presented in Figure~\ref{fig2}(a). Here the minor radius of the oblate spheroidal nanoparticle has been fixed to 55~nm. The absorption enhancement has been calculated as the power absorbed in 100-nm GaAs surface film with silver nanoantenna array normalized to the power absorbed in the GaAs film without nanoantennas (the power lost in the metal has been removed from these calculations). It is clearly seen that there are two strong local maxima of the absorption enhancement in the simulation map accompanied by few "hot spots" in the map. The first maximum is in the narrow range of the distances near 280~nm and in the wide range of radii from 70~nm to 115~nm. The second absorption maximum is at the distances of 650~--~730~nm for the particles with radii ranging between 85~nm and 105~nm. The latter is impractical for our case, since the sparse nanoantenna distribution will result in a very low number of nanoparticles inside the PCA gap ($\sim 100$ per gap), and inherent absence of the collective interactions. Thus, from these numerical calculations we may expect at least 4-fold enhancement of the THz photoconductive antenna performance with such optimized plasmonic nanostructures. Below we show that such Ag nanostructures may be fabricated via cost-effective method of thermal dewetting, which does not require any post processing. 

The electric field distribution profiles over the air-GaAs interface with and without optimized Ag spheroidal nanoantennas are shown in Figure~\ref{fig2}(b,c). The structure is excited by a normally incident plane electromagnetic wave with the wavelength of 800~nm (the wavelength of the fs-laser used in our setup, see \textit{Methods}). The geometrical parameters of nanoparticles correspond to the optimal values obtained from the above mentioned modeling results. The larger axis of the spheroid is D~=~168~nm, the minor axis of the spheroid is d~=~106~nm. The distance between nanoparticle centers is a~=~280~nm. It can be seen that the magnitude of the electric field under the spheroid (Figure~\ref{fig2}~(c)) is larger by a factor of 2.5 than the magnitude of the field at the interface between air and GaAs without nanoparticle (Figure~\ref{fig2}(b)). We would like to note that, while nanoparticles are on the surface of a high-index substrate (GaAs), the electric field is mostly concentrated in a thin layer of the substrate surface. The collective interactions excited in the array of nanoparticles cause additional increase of electric field in the gaps between them. These modes, localized and collective, enhance the absorption of fs-laser radiation near nanoantennas and may result in photocurrent density and, thus, THz emission enhancement. The power $P_{\rm abs}$, absorbed in a volume $V'$ of substrate depends on the electric field distribution inside the substrate and may be estimated as $P_{\rm abs}\sim|\boldsymbol{E}|^2V'$, where $\boldsymbol{E}$ is a electric field vector (averaged over the volume $V'$) in the medium~\cite{Lepeshov2016b}. Thus, the absorbed power is directly proportional to the value of $|\boldsymbol{E}|^2$ and, therefore, the strong localization of electric field in the semiconductor near the nanoparticles causes an increase of the absorbed power, which results in more effective photoexcitation of the free carriers. Figure~\ref{fig2}(d,e) shows a comparison of absorbed power densities in the semiconductor with and without nanoparticles. One has to note that the power absorbed in GaAs with silver nanoantenna is significantly higher than in pure GaAs near the surface.

\begin{figure*}[!t]
	\begin{center}
		\includegraphics[width=0.99\linewidth]{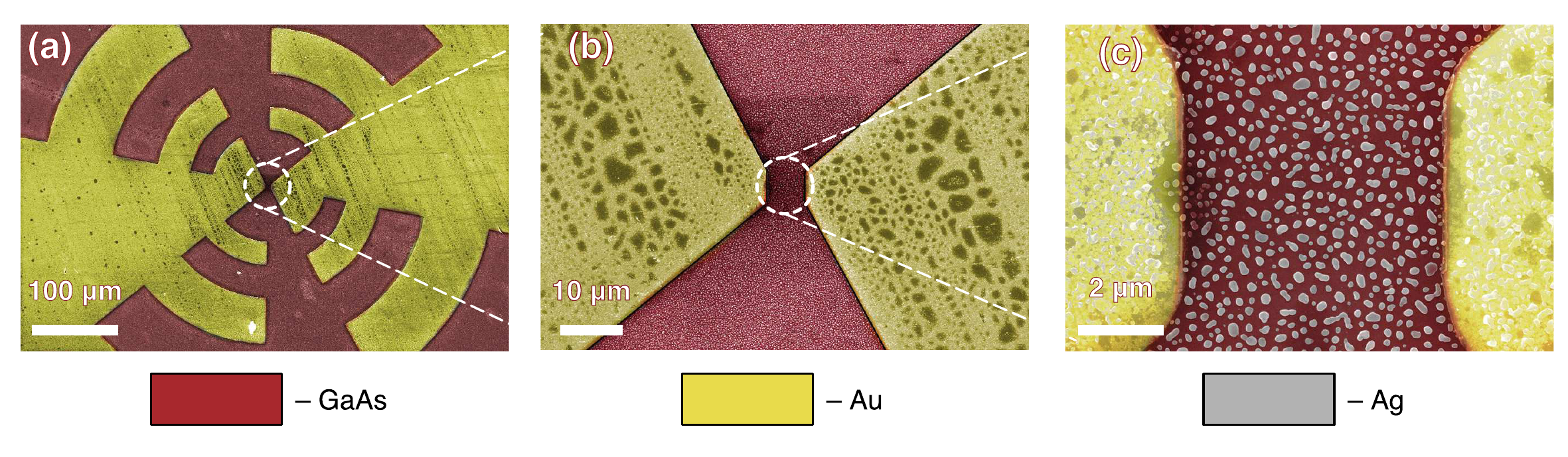}
	\end{center}
	\caption{The typical SEM images of the fabricated log-periodic THz PCA (a), its central part (b), and its gap filled with silver nanoantennas (c).}\label{fig0}
\end{figure*}

Next, to prove our theoretical findings, we perform experimental investigations of hybrid PCAs with optimized Ag nanoantennas and compare the obtained results with previous studies~\cite{Park2012, Jeong_ACSN_12}. The PCAs were fabricated on semi-insulating GaAs substrates containing self assembled InAs quantum-dots (QDs). Such QD based antennas have been recently demonstrated to generate effectively both pulsed~\cite{Leyman2016, Gorodetsky2016} and CW~\cite{Fedorova2016a} THz radiation. Since the 800~nm wavelength pump is used, and the carriers are generated in the whole volume of the GaAs matrix, the dots serve only as carrier lifetime shorteners~\cite{Gorodetsky2015}, in a similar way as defects in low temperature grown GaAs.  The log-periodic shape of electrodes has been selected for the realization of the THz antenna due to the fact that such design provides a broadband radiative spectrum. The latter circumstance allows studying the THz antenna operation in the wider-frequency range. The log-periodic antenna electrodes have been deposited onto the QD-based substrate by UV optical lithography. The fabricated THz antennas have a smallest gap of 8~$\mu$m, and overall diameter of 1.8~mm. Then, a 20-nm silver film has been deposited onto the substrate surface of one of the fabricated antennas. Upon thermal dewetting process caused by heating, the silver film has been transformed into a disordered array of spheroid nanoparticles. The size of the resulting nanoparticles depends on the thickness of the silver film~\cite{Gladskikh2015, Toropov2017} and can further be changed by laser pump~\cite{Vartanyan2016, Toropov2017}. The sizes of the fabricated nanoparticles proximately correspond to calculated optimal ones and the average distance between nanoparticles is 280~nm. The dashed oval in Figure~\ref{fig2}(a) shows the location of the resulting nanoantennas on the radius-center distance map. The SEM pictures of the produced hybrid THz antenna are shown in Figure~\ref{fig0}. For more details about the structures fabrication see \textit{Methods}. We would like to note that the nanoparticle array can be produced also by laser assisted methods of thermal dewetting~\cite{Makarov2016}.

\begin{figure*}[!t]
	\begin{center}
		\includegraphics[width=0.99\linewidth]{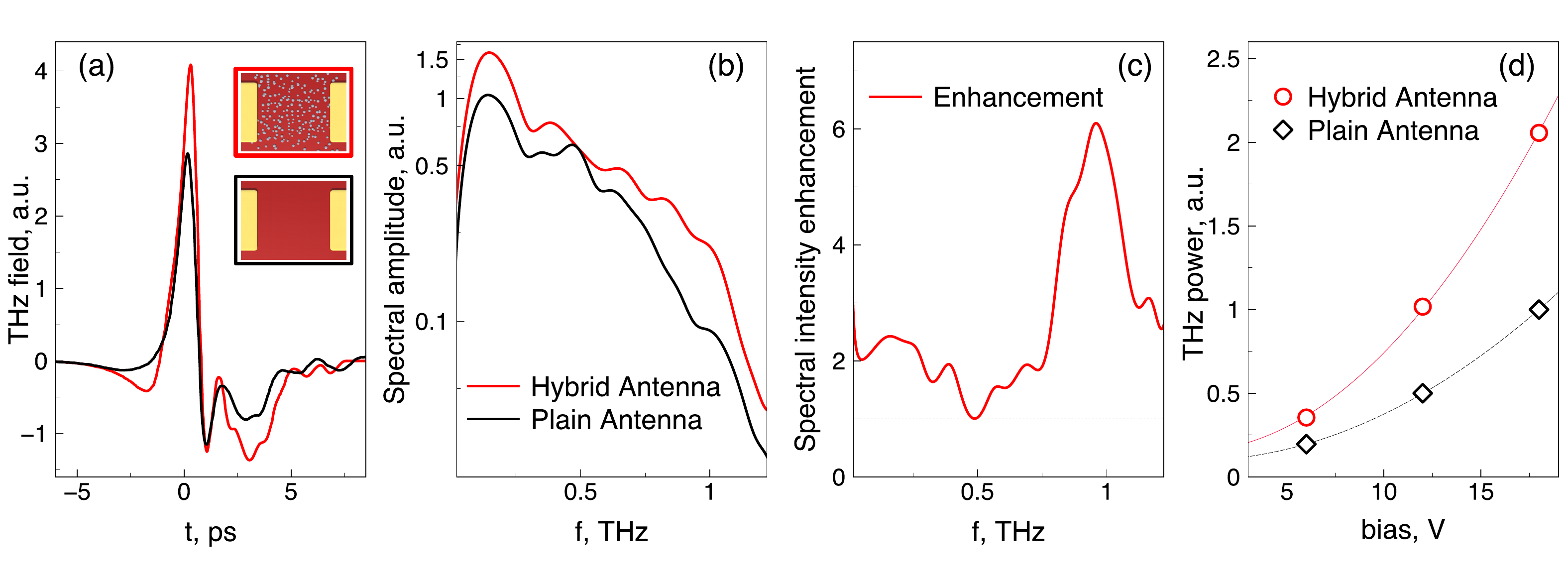}
	\end{center}
	\caption{Comparison of optimized design silver nanoantenna enhanced hybrid PCA performance with plain PCA. (a)~Time domain THz field profiles, and (b)~their corresponding spectral amplitudes. Plain antenna signal is plotted in black, hybrid antenna signal in red. (c)~Spectral intensity enhancement gained due to optimized silver nanoantennas in the gap of the PCA, (d)~power trends vs bias for optimized hybrid (red) and plain (black) PCAs. Markers correspond to experimentally obtained values, and curves are the $P\propto V^2$ fits.}\label{fig3}
\end{figure*}

For experimental verification of the proposed hybrid antennas, a standard THz time-domain spectroscopic (TDS) system has been used. THz-TDS is pumped with Sprite-XT (M Squared Ltd.) femtosecond Ti:sapphire laser that delivers pulses of 120~fs duration at 80~MHz repetition rate with central wavelength of 800~nm. As a THz detector, a LT-GaAs photoconductive antenna made by Teravil Ltd. was used, and the beam was guided between the transmitter and detector by two off-axis parabolic mirrors. To estimate the effect of silver nanoantennas, signals from standard and nanoantenna-enhanced THz antennas were measured for similar pumping and bias conditions. The experimental results are summarized in Figure~\ref{fig3}. 

Figure~\ref{fig3}(a) shows the time domain THz signals generated with the log-periodic PCA with and without Ag spheroidal nanoantennas for different bias voltage. The corresponding signal amplitude spectra, obtained using the Fourier transform, are shown in Figure~\ref{fig3}(b). It can be seen that the antenna demonstrates an uneven enhancement across the spectrum. The spectrum of THz signal intensity enhanced by optical nanoantennas is shown in Figure~\ref{fig3}(c). The enhancement reaches its maximum around 1~THz, and the enhancement value corresponds to the 5-fold theoretically derived value of the power absorption enhancement. Lower effect at other frequencies and negligible enhancement at 0.5~THz can be associated with the change in the THz antenna impedance induced by highly conductive silver in the gap. The overall power of generated THz signal with (red curve) and without (black curve) nanoantennas are plotted as a function of the PCA bias in Figure~\ref{fig3}(d). The results demonstrate significant amplification of the emitted THz signal spectral power with increasing of the bias voltage. The results for THz PCAs with nanoantennas demonstrate over 5-fold increase in comparison with the case of nanoantenna absent. This is 3 times higher than the results that have been previously reported~\cite{Park2012}.

In conclusion, we have demonstrated, both theoretically and experimentally, an unprecedented enhancement of the photoconductive antenna operation efficiency by optimized spheroidal Ag plasmonic nanoantennas. The resulting hybrid PCA demonstrates over 5-fold increase in the generated THz signal around 1~THz, and over 2-fold increase in overall generated THz power, which coincides with theoretical estimations. Moreover, we have proposed the cost-effective fabrication procedure to realize such hybrid THz antennas with optimized plasmonic nanostructures via thermal dewetting process, which does not require any post processing and makes the proposed solution very attractive for applications. We believe that our results may be useful for many relevant applications, requiring compact and effective room-temperature THz sources, including spectroscopy, biological sensing, security imaging, and ultrafast data transmission.

\section*{Methods}

The QD wafer of the PCA was grown by molecular beam epitaxy (MBE) in the Stranski-Krastanov regime on a GaAs substrate. The QDs were set in 40 layers, resulting in overall active region thickness of $\sim1$~$\mu$m. Each QD layer was grown by deposition of 2.3 monolayer (ML) InAs at $500^{\circ}$C, capped by 4~nm ${\rm In_{0.15}Ga_{0.85}As}$ then by 6~nm GaAs, after which the temperature was raised to $580^{\circ}$C before growth of the subsequent 30~nm GaAs spacer layer, in order to desorb segregated In. Also, prior to growth of each QD layer, the GaAs surface was annealed under an ${\rm As_2}$ flux for 5 minutes to flatten the growth surface. AFM analysis gives an estimated QD density of around $3\cdot10^{10}$~cm$^{-2}$ per layer. 

Metallic antenna electrodes were deposited over a semiconductor substrate using a standard UV photolithography and further wet etching of the surface Ni/Au (40~nm/200~nm, respectively) features. Post-process annealing to increase Ohmic contact between the antenna metal and GaAs surfaces was applied.

Plasmonic nanoparticles were prepared by thermal dewetting of a silver thin film. The Ag film has been fabricated by deposition of silver (99.99\% purity) at ultra-high vacuum conditions, residual pressure was of $\sim10^{-7}$~mbar. The thickness of the obtained silver film was controlled using the quartz crystal microbalances and was equivalent to 20~nm. To obtain a clearly resolved morphology, the samples with as-prepared antenna were kept at $250^{\circ}$C during the deposition process and 1 hour after that. According to scanning electron microscopy investigations, average lateral size in diameter of the islands was of 150~nm. Surface density was estimated to be $\sim10^9$~cm$^{-2}$. Such conditions of deposition and thermal annealing provide both coupling of quantum dots with plasmonic excitations and did not shunt the substrate with the THz antenna on its surface.

\begin{acknowledgements}
This work was supported by Russian Foundation for Basic Research (Project 16-07-01166~a)  and the Welch Foundation with grant No. F-1802. Authors thank M.Baranov for the SEM pictures.
\end{acknowledgements}

%\bibliography{THzConversion_V4}

%

\end{document}